\documentclass[12pt]{article}
\begin{document}
\title{The effect of annealing on the elastoplastic
and viscoelastic responses of isotactic polypropylene}

\author{Aleksey D. Drozdov and Jesper deClaville Christiansen\\
Department of Production\\
Aalborg University\\
Fibigerstraede 16\\
DK--9220 Aalborg, Denmark}
\date{}
\maketitle

\begin{abstract}
Observations are reported on isotactic polypropylene
(i) in a series of tensile tests with a constant strain rate 
on specimens annealed for 24 h at various temperatures 
in the range from 110 to 150 $^{\circ}$C and 
(ii) in two series of creep tests in the sub-yield region of deformation
on samples not subjected to thermal treatment 
and on specimens annealed at 140 $^{\circ}$C.
A model is developed for the elastoplastic and
nonlinear viscoelastic responses of semicrystalline polymers.
A polymer is treated an equivalent transient network of macromolecules 
bridged by junctions (physical cross-links, entanglements and lamellar blocks).
The network is assumed to be highly heterogeneous,
and it is thought of as an ensemble of meso-regions 
with different activation energies for separation of strands from 
temporary nodes.
The elastoplastic behavior is modelled as sliding of meso-domains
with respect to each other driven by mechanical factors.
The viscoelastic response is attributed to detachment of active strands 
from temporary junctions and attachment of dangling chains to the network.
Constitutive equations for isothermal uniaxial deformation
are derived by using the laws of thermodynamics.
Adjustable parameters in the stress--strain relations are found
by fitting the experimental data.
\end{abstract}

\section{Introduction}

This paper is concerned with the effect of annealing on the elastoplastic 
and viscoelastic responses of injection-molded isotactic polypropylene (iPP).
This semi-crystalline polypropylene is chosen for the analysis 
because of its numerous applications in industry 
(oriented films for packaging,
reinforcing fibres,
nonwoven fabrics,
blends with thermoplastic elastomers, etc.).

The effect of annealing at elevated temperatures on the morphology 
of polypropylene has been a focus of attention in past five years,
see, e.g., \cite{KB97,RQS98,XSM98,ABM99,IS00,MHY00,GHT02},
to mention a few.
The previous works concentrated on calorimetric studies
of morphological transformations driven by primary and
secondary crystallization.
The influence of changes in the micro-structure of spherulites
on the mechanical behavior of iPP was not studied in detail.

The nonlinear viscoelastic response of polypropylene was analyzed
by Ward and Wolfe \cite{WW66} and Smart and Williams \cite{SW72}
several decades ago and, in the past decade, 
by Wortmann and Schulz \cite{WS94,WS95},
Ariyama \cite{Ari96},
Ariyama et al. \cite{AMK97},
Dutta and Edward \cite{DE97},
Read and Tomlins \cite{RT97},
Tomlins and Read \cite{TR98}
and Sweeney et al. \cite{SCC99}.

Yielding and viscoplasticity of iPP have been recently investigated 
by Aboulfaraj et al. \cite{AGU95},
Kalay and Bevis \cite{KB97},
Coulon et al. \cite{CCG98},
Seguela et al. \cite{SeSE99},
Staniek et al. \cite{StSE99},
Nitta and Takayanagi \cite{NT99,NT00}
and Labour et al. \cite{LGS01},
to mention a few.

Isotactic polypropylene is a semi-crystalline polymer containing
monoclinic $\alpha$ crystallites,
hexagonal $\beta$ structures,
orthorhombic $\gamma$ polymorphs
and ``smectic" mesophase \cite{IS00}.
At rapid cooling of the melt (at the stage of injection molding),
$\alpha$ crystallites and smectic mesophase are mainly developed,
whereas $\beta$ and $\gamma$ polymorphs are observed as minority components
\cite{KB97,RQS98}.

A unique feature of $\alpha$ spherulites in iPP is the 
lamellar cross-hatching: 
development of transverse lamellae oriented in the direction 
perpendicular to the direction of radial lamellae \cite{IS00,MHY00}.
The characteristic size of $\alpha$ spherulites in injection-molded 
specimens is estimated as 100 to 200 $\mu$m \cite{KB97,CCG98}.
These spherulites consist of crystalline lamellae with thickness 
of 10 to 20 nm \cite{MHY00,CCG98}.

The amorphous phase is located  (i) between spherulites,
(ii) inside spherulites, in ``liquid pockets" between lamellar stacks \cite{VMH96},
and (iii) between lamellae in lamellar stacks.
It consists of (i) relatively mobile chains between spherulites, 
in liquid pockets and between radial lamellae inside lamellar stacks,
and (ii) severely restricted chains in the regions bounded 
by radial and tangential lamellae (rigid amorphous fraction  \cite{VMH96}).

Stretching of iPP specimens results in inter-lamellar separation,
rotation and twist of lamellae,
fine and coarse slip of lamellar blocks and
their fragmentation \cite{AGU95,SeSE99},
chain slip through the crystals, 
sliding, pull-out and breakage of tie chains \cite{NT99,NT00},
and activation of rigid amorphous fraction.
At large strains, these morphological transformations lead to
cavitation, formation of fibrills and stress-induced crystallization of iPP
\cite{ZBC99}.

To develop stress--strain relations, we apply a method 
of ``homogenization of micro-structure,"
according to which a sophisticated morphology of isotactic polypropylene
is modelled by an equivalent phase whose deformation captures essential 
features of the mechanical response.
An amorphous phase is chosen as the equivalent phase for the following reasons:
\begin{enumerate}
\item
The viscoelastic response of isotactic polypropylene
is conventionally associated with rearrangement of chains 
in amorphous regions \cite{CCG98}.

\item
Sliding of tie chains along and their detachment from lamellae 
play the key role in the time-dependent response of iPP \cite{NT99,NT00}.

\item
The viscoplastic flow in semi-crystalline polymers
is assumed to be ``initiated in the amorphous phase 
before transitioning into the crystalline phase" \cite{MP01}.

\item
The time-dependent behavior of polypropylene is conventionally modelled
within the concept of a network of macromolecules \cite{SCC99,SW96}.
\end{enumerate}

Dynamic mechanical analysis reveals that the loss tangent 
of iPP demonstrates two pronounced maxima 
being plotted versus temperature \cite{SeSE99,And99}.
The first maximum ($\beta$--transition in the interval 
between $T=-20$ and $T=10$ $^{\circ}$C) is associated with 
the glass transition in a mobile part of the amorphous phase, 
whereas the other maximum ($\alpha$--transition in the interval 
between $T=70$ and $T=110$ $^{\circ}$C) is attributed to the
glass transition in the remaining part of the amorphous phase,
the rigid amorphous fraction \cite{VMH96}.

At room temperature (i.e., above the glass transition temperature 
for the mobile amorphous phase) iPP is treated a transient network 
of macromolecules \cite{TE92} bridged by junctions (physical cross-links,
entanglements and lamellar blocks).
The network is assumed to be highly heterogeneous
(the inhomogeneity is attributed to interactions between spherulites
and amorphous regions, as well as to local density fluctuations
in the amorphous phase),
and it is thought of as an ensemble of meso-regions (MRs)
with different activation energies for separation of strands from 
temporary nodes.

Two types of MRs are distinguished: 
(i) active domains, where strands separate from junctions as 
they are thermally agitated (mobile part of the amorphous phase), 
and (ii) passive domains, where detachment of chains 
from junctions is prevented by surrounding lamellae.

Stretching of a specimen induces
\begin{enumerate}
\item
slippage of meso-domains with respect to each other 
(which models the elastoplastic behavior of iPP),

\item
separation of active strands from temporary junctions and
attachment of dangling chains to the network in active meso-regions
(which reflects the viscoelastic response),

\item
an increase in the content of active MRs driven by release 
of the rigid amorphous fraction due to lamellar fragmentation
(which is associated with the nonlinearity of the viscoelastic
behavior in the sub-yield region of deformation).
\end{enumerate}

The objective of this study is three-fold:
\begin{enumerate}
\item
To report experimental data in tensile tests with a constant strain
rate on specimens annealed at various temperatures in the interval 
between 110 and 150 $^{\circ}$C.

\item
To present obsevations in creep tests on non-annealed samples
and on specimens annealed at 140 $^{\circ}$C in the sub-yield region.

\item
To derive constitutive equations for the time-dependent behavior 
of a semicrystalline polymer and to find adjustable parameters
in the stress--strain relations by fitting observations.
\end{enumerate}
The goal of this paper is to shed some light on correlations between
morphological transformations in iPP at annealing
and changes in stress--strain diagrams and creep curves
measured in the sub-yield and post-yield regions of deformation.

The exposition is organized as follows.
The specimens and the experimental procedure are described
in Section 2.
Section 3 deals with kinetic equations for rearrangement of 
strands in active meso-domains.
Sliding of MRs with respect to each other is described in 
Section 4.
Stress--strain relations are derived in Section 5 by using the laws
of thermodynamics.
Adjustable parameters in the constitutive equations
are found in Section 6 by fitting observations.
A brief discussion of our findings is presented in Section 7.
Some concluding remarks are formulated in Section 8.

\section{Experimental procedure}

Isotactic polypropylene (Novolen 1100L) was supplied by BASF (Targor).
ASTM dumbbell specimens were injection molded 
with length 148 mm, width 10 mm and height 3.8 mm.
Mechanical tests were performed on non-annealed specimens,
as well as on samples annealed in an oven for 24 h at the
temperatures 110, 120, 130, 140 and 150 $^{\circ}$C.
After annealing, the specimens were slowly cooled by air.
To minimize the effect of physical aging on the time-dependent response,
tests were carried out a week after thermal pre-treatment.

Uniaxial tensile tests were performed 
at room temperature on a testing machine Instron--5568 
equipped with electro-mechanical sensors 
for the control of longitudinal strains in the active zone of samples 
(with the distance 50 mm between clips).
The tensile force was measured by a standard load cell.
The longitudinal stress, $\sigma$, was determined
as the ratio of the axial force to the cross-sectional area
of stress-free specimens.

In the first series of tests, non-annealed specimens and specimens annealed
at various temperatures, $T$,  were loaded with the cross-head speed 5 mm/min 
(that corresponded to the Hencky strain rate 
$\dot{\epsilon}_{H}=1.1\cdot 10^{-3}$ s$^{-1}$) up to the maximal strain $\epsilon_{\max}=0.06$.
According to \cite{ABJ95}, the chosen strain rate ensured
nearly isothermal test conditions.
The engineering stress, $\sigma$, was plotted versus the longitudinal strain,
$\epsilon$, in Figure 1.

A series of 6 creep tests was performed on non-annealed specimens
at the longitudinal stresses
$\sigma_{1}^{0}=10.0$ MPa,
$\sigma_{2}^{0}=15.0$ MPa,
$\sigma_{3}^{0}=17.5$ MPa,
$\sigma_{4}^{0}=20.0$ MPa,
$\sigma_{5}^{0}=22.5$ MPa
and
$\sigma_{6}^{0}=25.0$ MPa.
Each creep test was carried out on a new sample.
In the $m$th test ($m=1,\ldots,6$), a specimen was loaded 
with the cross-head speed 5 mm/min up to the engineering stress $\sigma_{m}^{0}$
that was preserved constant during the creep test, $t_{\rm c}=20$ min.
The longitudinal strain, $\epsilon$, was plotted versus 
the logarithm ($\log=\log_{10}$) of time $t$ 
(the instant $t=0$ corresponds to the beginning of a creep test) in Figure 2.

To evaluate the effect of annealing on the nonlinear viscoelastic behavior of iPP,
a series of 5 creep tests was performed on specimens annealed at 140 $^{\circ}$C
at the longitudinal stresses
$\sigma_{1}^{0}=10.0$ MPa,
$\sigma_{2}^{0}=15.0$ MPa,
$\sigma_{3}^{0}=20.0$ MPa,
$\sigma_{4}^{0}=25.0$ MPa,
and 
$\sigma_{5}^{0}=30.0$ MPa.
The strain, $\epsilon$, was plotted versus the logarithm of time, $t$, in Figure 3.

Figures 2 and 3 demonstrate that the rate of increase in strain, $\epsilon$, with time, $t$,
is relatively low at small stresses, and it noticeably grows with the longitudinal stress,
$\sigma$.
Our aim is to derive constitutive equations for the elastoplastic and viscoelastic
responses of a semi-crystalline polymer and to find adjustable parameters in the
stress--strain relations by fitting the experimental data plotted in Figures 1 to 3.

\section{Rearrangement of strands in active meso-regions}

A semicrystalline polymer is treated as a transient network of 
macromolecules bridged by temporary nodes.
The network is modelled as an ensemble of meso-regions with
various potential energies for detachment of strands from their
junctions.
Two types of meso-domains are distinguished: 
\begin{enumerate}
\item
passive, where all nodes are thought of as permanent,

\item
active, where active strands (whose ends are connected to 
contiguous junctions) separate from the nodes 
at random times when these strands are thermally agitated.
\end{enumerate}
An active chain whose end detaches from a junction is transformed
into a dangling chain.
A dangling chain  returns into the active state
when its free end captures a nearby junction at a random instant.

Denote by $X_{\rm a}$ the number (per unit mass) of active strands
in active MRs, and by $X_{\rm p}$ the number (per unit mass)
of strands connected to the network in passive MRs.
Under a time-dependent loading program,
some lamellae (restricting mobility of chains in passive MRs) break,
which results in an increase in the number of strands to be rearranged.
The quantities $X_{\rm a}$ and $X_{\rm p}$ are treated as functions 
of time, $t$, that obey the conservation law
\begin{equation}
X_{\rm a}(t) +X_{\rm p}(t) =X,
\end{equation}
where $X$ is the average number of active strands per unit mass
of a polymer (which is assumed to be time-independent).

Rearrangement of strands in active MRs is thought of as
a thermally activated process.
The rate of detachment of active strands from their junctions
in a MR with potential energy $\bar{\omega}$ is given by the Eyring equation \cite{KE75}
\[
\Gamma=\Gamma_{\rm a}\exp\biggl (-\frac{\bar{\omega}}{k_{\rm B}T}\biggr ),
\]
where $k_{\rm B}$ is Boltzmann's constant, 
$T$ is the absolute temperature, 
and the pre-factor $\Gamma_{\rm a}$ is independent
of energy $\bar{\omega}$ and temperature $T$.
Confining ourselves to isothermal deformations at
a reference temperature $T_{0}$ and introducing 
the dimensionless activation energy
$\omega=\bar{\omega}/(k_{\rm B}T_{0})$, 
we arrive at the formula
\begin{equation}
\Gamma=\Gamma_{\rm a}\exp (-\omega).
\end{equation}
An ensemble of active MRs with various potential 
energies is described by the distribution function $p(t,\omega)$ that equals the
ratio of the number, $N_{\rm a}(t,\omega)$, of active meso-domains 
with energy $\omega$ at instant $t$ to the total number of active MRs,
\begin{equation}
p(t,\omega)=\frac{N_{\rm a}(t,\omega)}{X_{\rm a}(t)},
\qquad
X_{\rm a}(t)=\int_{0}^{\infty} N_{\rm a}(t,\omega) d\omega,
\end{equation}
and by  the concentration of active MRs
\begin{equation}
\kappa_{\rm a}(t)=\frac{X_{\rm a}(t)}{X}.
\end{equation}
In what follows, constitutive equations will be derived
for an arbitrary distribution function $p(t,\omega)$.
To fit experimental data, a random energy model is applied with
\begin{equation}
p(t,\omega) = p_{0}(t) \exp \biggl [ -\frac{(\omega-\Omega(t))^{2}}{2\Sigma^{2}(t)} \biggr ]
\quad (\omega\geq 0), \qquad
p(t,\omega)=0 \quad (\omega <0),
\end{equation}
where $\Omega$ is the average activation energy in an ensemble of
active meso-domains,
$\Sigma$ is the standard deviation of potential energies for separation of
strands,
and $p_{0}(t)$ is determined by the condition
\begin{equation}
\int_{0}^{\infty} p(t,\omega) d\omega =1.
\end{equation}
An ensemble of active meso-domains is characterized by the
function $n_{\rm a}(t,\tau,\omega)$ that equals the number 
(per unit mass) of active strands at time $t$ belonging to
active MRs with potential energy $\omega$ that have last been 
rearranged before instant $\tau\in [0,t]$.
In particular, $n_{\rm a}(t,t,\omega)$ is the number (per unit mass)
of active strands in active MRs with potential 
energy $\omega$ at time $t\geq 0$,
\begin{equation}
n_{\rm a}(t,t,\omega)=N_{\rm a}(t,\omega). 
\end{equation}
The amount $\varphi(\tau,\omega)d\tau$, where
\begin{equation} 
\varphi(\tau,\omega)= \frac{\partial n_{\rm a}}{\partial \tau}(t,\tau,\omega)\biggl |_{t=\tau}, 
\end{equation}
equals the number (per unit mass) of dangling strands in 
active MRs with potential energy $\omega$  that merge 
with the network within the interval $[\tau,\tau+d\tau ]$,
and the quantity
\[ \frac{\partial n_{\rm a}}{\partial \tau}(t,\tau,\omega) d\tau \]
is the number of these strands that have not detached from
temporary junctions during the interval $[\tau, t]$.
The number (per unit mass) of strands in active MRs that separate (for
the first time) from the network within the interval $[t,t+dt]$ reads
\[ -\frac{\partial n_{\rm a}}{\partial t}(t,0,\omega) dt. \]
The number (per unit mass) of strands in active MRs 
that merged with the network during the interval $[\tau,\tau+d\tau ]$
and, afterwards, separated from the network within the interval $[t,t+dt]$ 
is given by
\[ -\frac{\partial^{2} n_{\rm a}}{\partial t\partial \tau}(t,\tau,\omega) dt d\tau. \]
The rate of detachment, $\Gamma$, equals the ratio of
the number of active strands that separate from the network per unit
time to the current number of active strands.
Applying this definition to active strands that merged with the network
during the interval $[\tau,\tau+d\tau ]$
and separated from temporary junctions within the interval $[t,t+dt]$, 
we find that
\begin{equation}
\frac{\partial^{2} n_{\rm a}}{\partial t\partial \tau}(t,\tau,\omega)=-
\Gamma(\omega) \frac{\partial n_{\rm a}}{\partial \tau}(t,\tau,\omega).
\end{equation}
Changes in the function $n_{\rm a}(t,0,\omega)$ are governed
by two processes: 
\begin{enumerate}
\item
detachment of active strands from temporary nodes,

\item
mechanically-induced activation of passive MRs.
\end{enumerate}
The kinetic equation for this function reads
\begin{equation}
\frac{\partial n_{\rm a}}{\partial t}(t,0,\omega)=-
\Gamma(\omega) n_{\rm a}(t,0,\omega)
+\frac{\partial N_{\rm a}}{\partial t}(t,\omega).
\end{equation}
The solution of Eq. (10) with initial condition (7), where we set $t=0$, 
is given by
\begin{equation}
n_{\rm a}(t,0,\omega) = N_{\rm a}(0,\omega)\exp \Bigl [ - \Gamma(\omega)t \Bigr ]
+\int_{0}^{t} \frac{\partial N_{\rm a}}{\partial t}(\tau,\omega) 
\exp \Bigl [ - \Gamma(\omega)(t-\tau) \Bigr ] d\tau .
\end{equation}
It follows from Eqs. (8) and (9) that 
\begin{equation}
\frac{\partial n_{\rm a}}{\partial \tau}(t,\tau,\omega)
=\varphi(\tau,\omega) \exp \Bigl [ - \Gamma(\omega)(t-\tau) \Bigr ] .
\end{equation}
Equation (7) implies that
\begin{equation}
N_{\rm a}(t,\omega) = n_{\rm a}(t,0,\omega)
+\int_{0}^{t} \frac{\partial n_{\rm a}}{\partial \tau}(t,\tau,\omega) d\tau.
\end{equation}
Differentiating Eq. (13) with respect to time and using Eq. (8), we find that
\[
\varphi(t,\omega)+\frac{\partial n_{\rm a}}{\partial t}(t,0,\omega)
+\int_{0}^{t} \frac{\partial^{2} n_{\rm a}}{\partial t\partial \tau}(t,\tau,\omega)
d\tau=\frac{\partial N_{\rm a}}{\partial t}(t,\omega).
\]
This equality together with Eqs. (9), (10) and (13) results in 
\[
\varphi(t,\omega) =  \Gamma(\omega)N_{\rm a}(t,\omega).
\]
This expression together with Eq. (12) yields
\begin{equation}
\frac{\partial n_{\rm a}}{\partial \tau}(t,\tau,\omega)
=\Gamma(\omega)N_{\rm a}(t,\omega)
\exp \Bigl [ -\Gamma(\omega)(t-\tau) \Bigr ].
\end{equation}
Rearrangement of strands in active MRs is described 
by Eqs. (2), (11) and (14).
Separation of active strands from their junctions
and detachment of dangling chains to the network
reflect the viscoelastic response of a semi-crystalline polymer.

\section{Sliding of meso-regions}

It is assumed that meso-domains are not rigidly connected,
but can slide with respect to each other under straining.
Sliding of meso-domains is treated as a rate-independent process
and is associated with the elastoplastic behavior of a semi-crystalline
polymer.
We suppose that an increase in strain, $\epsilon$, by an increment,
$d\epsilon$, causes growth of the elastoplastic strain, $\epsilon_{\rm p}$,
by an increment, $d\epsilon_{\rm p}$, that is proportional to $d\epsilon$,
\[
d\epsilon_{\rm p}=\Psi d\epsilon.
\]
The coefficient of proportionality, $\Psi$, depends, in general, 
on the macro-strain, $\epsilon$, the macro-stress, $\sigma$, 
and the elastoplastic strain, $\epsilon_{\rm p}$.
We presume that $\Psi$ is a function of the elastic strain, $\epsilon_{\rm e}$, which
is defined as the difference between the macro-strain, $\epsilon$, and the
elastoplastic strain, $\epsilon_{\rm p}$,
\begin{equation}
\epsilon_{\rm e}(t)=\epsilon(t)-\epsilon_{\rm p}(t).
\end{equation}
This results in the kinetic equation
\begin{equation}
\frac{d\epsilon_{\rm p}}{dt}(t)=\Psi \Bigl (\epsilon(t)-\epsilon_{\rm p}(t) \Bigr ) 
\frac{d\epsilon}{dt}(t),
\qquad
\epsilon_{\rm p}(0)=0.
\end{equation}
It is natural to suppose that the function $\Psi(\epsilon_{\rm e})$ vanishes 
at $\epsilon_{\rm e}=0$ (no elastoplastic strains are observed at small macro-strains),
monotonically increases with the elastic strain,
and reaches some limiting value $b\in (0,1)$ at large values of $\epsilon_{\rm e}$
(which corresponds to a steady regime of plastic flow).
To minimize the number of adjustable parameters in constitutive equations,
an exponential dependence is adopted,
\begin{equation}
\Phi(\epsilon_{\rm e})=b\Bigl [ 1-\exp(-a \epsilon_{\rm e})\Bigr ],
\end{equation}
where the positive coefficients $a$ and $b$ are found by matching observations.
It is worth noting that Eqs. (16) and (17) differ from conventional flow rules 
in elastoplasticity, where the elastoplastic strain is assumed 
to be proportional to stress, $\sigma$.
These equations are, however, in good agreement with the conclusion by
Men and Strobl \cite{MS02} that ``tensile deformations of semi-crystalline
polymers \ldots are strain-controlled."

\section{Constitutive equations}

An active strand is modelled as a linear elastic medium with the mechanical energy
\begin{equation}
w(t)=\frac{1}{2}\mu e^{2}(t), 
\end{equation}
where $\mu$ is the average rigidity per strand
and $e$ is the strain from the stress-free state to the deformed state of the strand.

For strands belonging to passive meso-domains, the strain $e$
coincides with the elastic strain $\epsilon_{\rm e}$.
Multiplying the strain energy per strand, Eq. (18), by the number of strands in
passive MRs, we find the mechanical energy of meso-domains 
where rearrangement of chains is prevented by surrounding lamellae,
\begin{equation}
W_{\rm p}(t)=\frac{1}{2}\mu X_{\rm p}(t)\epsilon_{\rm e}^{2}(t). 
\end{equation}
With reference to the conventional theory of temporary networks \cite{TE92},
stresses in dangling strands are assumed to totally relax before
these strands merge with the network.
This implies that the reference (stress-free) state of a strand that
is attached to the network at time $\tau$ coincides with
the deformed state of the network at that instant.
For active strands that have not been rearranged until time $t$,
the strain $e(t)$ coincides with $\epsilon_{\rm e}(t)$, 
whereas for active strands that have last been merged with the network 
at time $\tau\in [0,t]$, the strain $e(t,\tau)$ is given by
\[ 
e(t,\tau)=\epsilon_{\rm e}(t)-\epsilon_{\rm e}(\tau).
\]
Summing the mechanical energies of active strands 
belonging to active MRs with various potential energies, $\omega$,
that were rearranged at various instants, $\tau\in [0,t]$, we find the
strain energy of active meso-domains,
\begin{equation}
W_{\rm a}(t) = \frac{1}{2}\mu \int_{0}^{\infty} 
\biggl \{ n_{\rm a}(t,0,\omega)\epsilon_{\rm e}^{2}(t)
+\int_{0}^{t} \frac{\partial n_{\rm a}}{\partial \tau}(t,\tau,\omega)
\Bigl [ \epsilon_{\rm e}(t)-\epsilon_{\rm e}(\tau)\Bigr ]^{2} d\tau \biggr \} d\omega .
\end{equation}
The mechanical energy per unit mass of a polymer reads
\[
W(t)=W_{\rm a}(t)+W_{\rm p}(t).
\]
It follows from this equality and Eqs. (15), (19) and (20) that
\begin{eqnarray}
W(t) &=&  \frac{1}{2}\mu \biggl \{  X_{\rm p}(t)
\Bigl (\epsilon(t)-\epsilon_{\rm p}(t) \Bigr )^{2}
+  \int_{0}^{\infty} \biggl [ n_{\rm a}(t,0,\omega)
\Bigl (\epsilon(t)-\epsilon_{\rm p}(t) \Bigr )^{2}
\nonumber\\
&& +\int_{0}^{t} \frac{\partial n_{\rm a}}{\partial \tau}(t,\tau,\omega)
\Bigl ( \Bigl ( \epsilon(t)-\epsilon_{\rm p}(t)\Bigr )
-\Bigl ( \epsilon(\tau)-\epsilon_{\rm p}(\tau)\Bigr ) \Bigr )^{2} d\tau \biggr ] d\omega 
\biggr \}.
\end{eqnarray}
Our aim now is to calculate the derivative of the function $W(t)$ with respect to time $t$.
Differentiation of Eq. (21) results in
\begin{equation}
\frac{dW}{dt}(t)=A(t) \Bigl [ \frac{d\epsilon}{dt}(t)-\frac{d\epsilon_{\rm p}}{dt}(t) \Bigr ]-A_{1}(t),
\end{equation}
where
\begin{eqnarray}
A(t) &=& \mu \biggl \{ X_{\rm p}(t) \Bigl ( \epsilon(t)-\epsilon_{\rm p}(t)\Bigr )
+\int_{0}^{\infty} \biggl [ n_{\rm a}(t,0,\omega) \Bigl ( \epsilon(t)-\epsilon_{\rm p}(t)\Bigr )
\nonumber\\
&& +\int_{0}^{t} \frac{\partial n_{\rm a}}{\partial \tau}(t,\tau,\omega)
\Bigl ( \Bigl ( \epsilon(t)-\epsilon_{\rm p}(t)\Bigr )
-\Bigl (\epsilon(\tau)-\epsilon_{\rm p}(\tau)\Bigr ) \Bigr )d\tau \biggr ]d\omega \biggr \},
\\
A_{1}(t) &=& -\frac{1}{2}\mu \biggl \{ \frac{d X_{\rm p}}{dt} (t)
\Bigl ( \epsilon(t)-\epsilon_{\rm p}(t) \Bigr )^{2}
+\int_{0}^{\infty} \biggl [ \frac{\partial n_{\rm a}}{\partial t}(t,0,\omega)
\Bigl ( \epsilon(t)-\epsilon_{\rm p}(t)\Bigr )^{2}
\nonumber\\
&& +\int_{0}^{t} \frac{\partial^{2} n_{\rm a}}{\partial t\partial \tau}(t,\tau,\omega)
\Bigl ( \Bigl (\epsilon(t)-\epsilon_{\rm p}(t)\Bigr )
-\Bigl (\epsilon(\tau)-\epsilon_{\rm p}(\tau)\Bigr )\Bigr )^{2} d\tau 
\biggr ]d\omega \biggr \}.
\end{eqnarray}
It follows from Eqs. (1), (3), (13) and (23) that
\begin{equation}
A(t) = \mu \biggl [ X \Bigl ( \epsilon(t)-\epsilon_{\rm p}(t)\Bigr )
-\int_{0}^{\infty} d\omega 
\int_{0}^{t} \frac{\partial n_{\rm a}}{\partial \tau}(t,\tau,\omega)
\Bigl ( \epsilon(\tau)-\epsilon_{\rm p}(\tau) \Bigr )d\tau \biggr ].
\end{equation}
Substituting Eqs. (9) and (10) into Eq. (24) and using Eqs. (1) and (3),
we obtain
\begin{eqnarray}
A_{1}(t) &=& \frac{1}{2}\mu \int_{0}^{\infty} \Gamma(\omega) d\omega
\biggl [ n_{\rm a}(t,0,\omega) \Bigl ( \epsilon(t)-\epsilon_{\rm p}(t)\Bigr )^{2}
\nonumber\\
&& +\int_{0}^{t} \frac{\partial n_{\rm a}}{\partial \tau}(t,\tau,\omega)
\Bigl ( \Bigl (\epsilon(t)-\epsilon_{\rm p}(t)\Bigr )
-\Bigl (\epsilon(\tau)-\epsilon_{\rm p}(\tau)\Bigr )\Bigr )^{2} d\tau \biggr ].
\end{eqnarray}
For isothermal uniaxial deformation, the Clausius-Duhem inequality reads
\[
Q(t)=-\frac{dW}{dt}(t)+\frac{1}{\rho}\sigma(t)\frac{d\epsilon}{dt}(t) \geq 0,
\]
where $\rho$ is mass density,
and $Q$ is internal dissipation per unit mass.
Substitution of Eqs. (16) and (22) into this equation results in
\begin{equation}
\biggl [ \sigma(t)-\rho A(t) \Bigl ( 1-\Psi(\epsilon(t)-\epsilon_{\rm p}(t)) \Bigr ) \biggr ]
\frac{d\epsilon}{dt}(t) + \rho A_{1}(t) \geq 0.
\end{equation}
It follows from Eq. (26) that the function $A_{1}(t)$ is non-negative for
an arbitrary program of loading.
This means that the dissipation inequality (27) is satisfied, provided that
the expression in the square brackets vanishes.
This condition together with Eq. (25) results in the stress--strain relation
\begin{eqnarray*}
\sigma(t)  &=& E \Bigl [ 1-\Psi(\epsilon(t)-\epsilon_{\rm p}(t)) \Bigr ]
\biggl [ \Bigl ( \epsilon(t)-\epsilon_{\rm p}(t)\Bigr )
\nonumber\\
&& -\frac{1}{X}\int_{0}^{\infty} d\omega 
\int_{0}^{t} \frac{\partial n_{\rm a}}{\partial \tau}(t,\tau,\omega)
\Bigl (\epsilon(\tau)-\epsilon_{\rm p}(\tau) \Bigr ) d\tau \biggr ],
\end{eqnarray*}
where $E=\rho\mu X$ is an analog of Young's modulus.
Substitution of Eqs. (3), (4) and (14) into this equality implies that
\begin{eqnarray}
\sigma(t) &=&  E \Bigl [ 1-\Psi(\epsilon(t)-\epsilon_{\rm p}(t)) \Bigr ]
\biggl [ \Bigl ( \epsilon(t)-\epsilon_{\rm p}(t)\Bigr )
-\kappa_{\rm a}(t) \int_{0}^{\infty} \Gamma(\omega) d\omega
\nonumber\\
&& \times \int_{0}^{t} \exp \Bigl (-\Gamma(\omega) (t-\tau)\Bigr )
\Bigl (\epsilon(\tau)-\epsilon_{\rm p}(\tau) \Bigr ) p(\tau, \omega) d\tau \biggr ].
\end{eqnarray}
Given functions $p(t,\omega)$ and $\kappa_{\rm a}(t)$,
Eqs. (2), (16), (17) and (28) determine the time-dependent behavior
of a semicrystalline polymer at isothermal uniaxial deformation.

To approximate the experimental data reported in Section 2,
we concentrate on tensile tests with constant strain rates
and on creep tests.

Confining ourselves to ``rapid" tensile tests, when the effect of
the material's viscosity on the mechanical response may be
disregarded, we neglect the integral term in Eq. (28) and arrive
at the constitutive equation
\begin{equation}
\sigma(t) =  E \Bigl [ 1-\Psi(\epsilon(t)-\epsilon_{\rm p}(t)) \Bigr ]
\Bigl ( \epsilon(t)-\epsilon_{\rm p}(t)\Bigr ).
\end{equation}
Equations (16), (17) and (29) are determined by 3 material parameters:
\begin{enumerate}
\item
the elastic modulus $E$,

\item
the constant $a$ that characterizes the rate of elastoplastic strain,

\item
the constant $b$ that describes a developed plastic flow.
\end{enumerate}
It is natural to assume that in a standard creep test with a longitudinal stress $\sigma^{0}$,
\[
\sigma(t)=\left \{\begin{array}{ll}
0,  & t<0,\\
\sigma^{0}, & t\geq 0,
\end{array} \right .
\]
the quantities $\kappa_{\rm a}$ and $p$ are functions of the stress intensity $\sigma^{0}$.
It follows from Eqs. (15) and (28) that the elastic strain, $\epsilon_{\rm e}$, reads
\begin{equation}
\epsilon_{\rm e}(t)=\frac{\sigma^{0}}{1-\Psi(\epsilon_{\rm e}(t))}
+\kappa_{\rm a}(\sigma^{0})    \int_{0}^{\infty} Z(t,\omega)p(\sigma^{0},\omega)d\omega,
\end{equation}
where 
\begin{equation}
Z(t,\omega )=\Gamma(\omega) \int_{0}^{t} \exp \Bigl [-\Gamma(\omega)(t-\tau)\Bigr ]
\epsilon_{\rm e}(\tau) d\tau.
\end{equation}
Equation (31) implies that the function $Z(t,\omega)$ satisfies the differential
equation
\begin{equation}
\frac{\partial Z}{\partial t}(t,\omega)=\Gamma(\omega) \Bigl [\epsilon_{\rm e}(t)-Z(t,\omega)\Bigr ],
\qquad
Z(0,\omega)=0.
\end{equation}
After solving Eqs. (30) and (32), the longitudinal strain, $\epsilon$, is found from Eqs. (15)
and (16) which can be presented in the form
\begin{equation}
\frac{d\epsilon}{dt}(t)=\Bigl [ 1-\Psi(\epsilon_{\rm e}(t))\Bigr ]^{-1} \frac{d\epsilon_{\rm e}}{dt}(t),
\qquad
\epsilon(0)=\epsilon^{0}(\sigma^{0}),
\qquad
\epsilon_{\rm e}(0)=\epsilon_{\rm e}^{0}(\sigma^{0}),
\end{equation}
where the initial conditions, $\epsilon^{0}(\sigma^{0})$ and $\epsilon_{\rm e}^{0}(\sigma^{0})$,
are determined by integration of Eqs. (16), (17) and (29) from $\sigma=0$ to $\sigma=\sigma^{0}$.

Given a stress $\sigma^{0}$, Eqs. (2), (5), (30), (32) and (33) are characterized 
by 4 adjustable parameters:
\begin{enumerate}
\item 
the average potential energy for separation of active strands $\Omega$,

\item
the standard deviation of potential energies of active MRs $\Sigma$,

\item
the concentration of active meso-domains $\kappa_{\rm a}$,

\item
the attempt rate for detachment of strands from temporary junctions
$\Gamma_{\rm a}$.
\end{enumerate}
It follows from Eqs. (2), (5), (30) and (32) that the quantities $\Omega$ 
and $\Gamma_{\rm a}$ are mutually dependent: an increase in $\Omega$
results in an increase in $\Gamma_{\rm a}$. 
To reduce the number of material constants to be found by matching
observations in creep tests, we set $\Gamma_{\rm a}=1$ s.
Our purpose now is to find adjustable parameters $E$, $a$, $b$, $\kappa_{\rm a}$,
$\Omega$ and $\Sigma$ by fitting experimental data
depicted in Figures 1 to 3.

\section{Fitting of observations}

We begin with the approximation of the stress--strain curves presented
in Figure 1.
Under uniaxial tension with the cross-head speed 5 mm/min, 
the strain $\epsilon_{\max}=0.06$ is reached within 69 s. 
According to Figures 2 and 3, changes in strain induced by rearrangement
of active strands during this period are insignificant at stresses up to 20 MPa, 
whereas the duration of stretching at higher stresses does not exceeds 30 s,
which causes rather small growth of strains.
Based on these observations, we treat the deformation process as rapid 
and apply Eqs. (16), (17) and (29) to fit experimental data.

For any temperature of annealing, $T$, the stress--strain curve
is matched independently.
To find the constants $E$, $a$ and $b$, we fix the intervals 
$[0,a_{\max}]$ and $[0,b_{\max}]$, 
where the ``best-fit" parameters $a$ and $b$ are assumed to be located,
and divide these intervals into $J$ subintervals by
the points $a_{i}=i\Delta a$ and $b_{j}=j\Delta b$  ($i,j=1,\ldots,J$)
with $\Delta a=a_{\max}/J$ and $\Delta b=b_{\max}/J$.
For any pair, $\{ a_{i}, b_{j} \}$, we integrate the governing equations
numerically (with the step $\Delta \epsilon=5.0\cdot 10^{-5}$)
by the Runge--Kutta method.
The elastic modulus $E=E(i,j)$ is found by the least-squares 
algorithm from the condition of minimum of the function
\[
K(i,j)=\sum_{\epsilon_{m}} \Bigl [ \sigma_{\rm exp}(\epsilon_{m})
-\sigma_{\rm num}(\epsilon_{m}) \Bigr ]^{2},
\]
where the sum is calculated over all experimental points, $\epsilon_{m}$,
depicted in Figure 1, $\sigma_{\rm exp}$ is the longitudinal stress 
measured in a tensile test, 
and $\sigma_{\rm num}$ is given by Eq. (29).
The ``best-fit" parameters $a$ and $b$ minimize 
$K$ on the set $ \{ a_{i}, b_{j} \quad (i,j=1,\ldots, J)  \}$.
After determining their values, $a_{i}$ and $b_{j}$, 
this procedure is repeated twice for the new intervals $[ a_{i-1}, a_{i+1}]$
and $[ b_{j-1}, b_{j+1}]$ to ensure an acceptable accuracy of fitting.

The ``best-fit" parameters $E$, $\epsilon_{\ast}=a^{-1}$ and $b$ are plotted in Figures 4 to 6
as functions of the annealing temperature $T$.
The experimental data are approximated by the linear functions
\begin{equation}
E=E_{0}+E_{1}T,
\qquad
\epsilon_{\ast}=\epsilon_{\ast 0}+\epsilon_{\ast 1}T,
\qquad
b=b_{0}+b_{1}T,
\end{equation}
where the coefficients $E_{m}$, $\epsilon_{\ast m}$ and $b_{m}$ $(m=0,1)$ are determined
by the least-squares technique.

We proceed with fitting the creep curves for non-annealed specimens
depicted in Figure 2.
For any stress, $\sigma^{0}$, the quantities $\epsilon^{0}$ and $\epsilon_{\rm ep}^{0}$
are found by integration of Eqs. (16), (17) and (29) with the material constants found in the
approximation of the stress--strain curve plotted in Figure 1.
To determine $\Omega(\sigma^{0})$, $\Sigma (\sigma^{0})$ and $\kappa(\sigma^{0})$,
the following algorithm is employed.
We fix the intervals $[0,\Omega_{\max}]$, $[0,\Sigma_{\max}]$ and $[0,\kappa_{\max}]$,
where the ``best-fit" parameters $\Omega$, $\Sigma$ and $\kappa$ are assumed to be located,
and divide these intervals into $J$ subintervals by
the points $\Omega_{i}=i \Delta \Omega$, $\Sigma_{j}=j\Delta \Sigma$ 
and $\kappa_{k}=k\Delta \kappa$ ($i,j,k=1,\ldots,J$)
with $\Delta \Omega=\Omega_{\max}/J$, $\Delta \Sigma=\Sigma_{\max}/J$
and $\Delta \kappa=\kappa_{\max}/J$.
For any pair, $\{ \Omega_{i}, \Sigma_{j} \}$, the constant $p^{0}=p^{0}(i,j)$
is found from Eq. (6), where the integral is evaluated 
by Simpson's method with 200 points and the step $\Delta \omega=0.15$.
For any triple, $\{ \Omega_{i}, \Sigma_{j}, \kappa_{k} \}$, 
Eqs. (2), (5), (30), (32) and (33) are integrated numerically 
(with the time step $\Delta t=0.1$) by the Runge--Kutta method.
The ``best-fit" parameters $\Omega$, $\Sigma$ and $\kappa$
minimize the function
\[
K(i,j,k)=\sum_{t_{m}} \Bigl [ \epsilon_{\rm exp}(t_{m})-\epsilon_{\rm num}(t_{m}) \Bigr ]^{2},
\]
where the sum is calculated over all experimental points, $t_{m}$,
presented in Figure 2, 
$\epsilon_{\rm exp}$ is the strain measured in the creep test, 
and $\epsilon_{\rm num}$ is given by Eq. (33).
After determining the ``best-fit" values, $\Omega_{i}$, $\Sigma_{j}$ and $\kappa_{k}$, 
this procedure is repeated for the new intervals
$[ \Omega_{i-1}, \Omega_{i+1}]$,
$[ \Sigma_{j-1}, \Sigma_{j+1}]$ and $[ \kappa_{k-1}, \kappa_{k+1}]$,
to provide an acceptable accuracy of fitting.

The adjustable parameters $\Omega$, $\Sigma$ and $\kappa$ are plotted versus
the engineering stress, $\sigma$, in Figures 7 to 9.
The experimental data are approximated by the linear functions
\begin{equation}
\Omega=\Omega_{0}+\Omega_{1}\sigma,
\qquad
\Sigma=\Sigma_{0}+\Sigma_{1}\sigma,
\qquad
\kappa=\kappa_{0}+\kappa_{1}\sigma,
\end{equation}
where the coefficients $\Omega_{m}$, $\Sigma_{m}$ and $\kappa_{m}$ ($m=0,1$)
are found by the least-squares technique.

Finally, we approximate the experimental data in creep tests on specimens
annealed for 24 h at $T=140$ $^{\circ}$C presented in Figure 3.
To determine adjustable parameters in the constitutive equations, 
the same procedure of fitting is used as for the observations depicted in Figure 2.
The ``best-fit" quantities $\Omega$, $\Sigma$ and $\kappa$ are plotted
versus the stress $\sigma$ in Figures 7 to 9 together with their approximations
by Eqs. (35).

\section{Discussion}

Figure 1 demonstrates fair agreement between the observations
in tensile tests with a constant strain rate 
and the results of numerical simulation.

Figure 4 shows that the elastic modulus $E$ monotonically increases
with annealing temperature $T$.
In the region of temperatures from room temperature to 130 $^{\circ}$C, 
this increase is rather small, 
but the rate of growth in $E(T)$ substantially increases in the interval
of temperatures between 130 and 150 $^{\circ}$C.
This conclusion is in agreement with observations in calorimetric tests
by other researchers, which reveal that the melting temperature, $T_{\rm m}$,
monotonically grows with crystallization temperature, $T$,
and the slope of the curve $T_{\rm m}(T)$ noticeably increases
at the temperature $T=130$ $^{\circ}$C.
Results of three DSC (differential scanning calorimetry)
studies on iPP are presented in Figure 10.
In this figure, the dependence of the melting peak, $T_{\rm m}$, 
on the crystallization temperature, $T$, is aproximated by the linear function
\begin{equation}
T_{\rm m}=c_{0}+c_{1}T.
\end{equation}
The coefficients $c_{m}$ ($m=0,1$) in Eq. (36) are found by using 
the least-squares algorithm.

According to the Gibbs--Thomson theory (see, e.g., \cite{IS00,GHT02}), 
an increase in the equilibrium melting temperature, $T_{\rm m}$, 
is tantamount to an increase in the average lamellar thickness.
Comparing Figures 4 and 10, we draw a conclusion that the growth
of the elastic modulus of iPP with annealing temperature may be attributed
to lamellar thickening.
 
Figures 5 and 6 reveal that the plastic strain, $\epsilon_{\ast}$,
and the rate of developed plastic flow, $b$, 
decrease in the interval from 110 to 130 $^{\circ}$C 
and increase at higher temperatures.
Two possible explanations may be provided for the decrease
in $\epsilon_{\ast}$:
\begin{enumerate}
\item
According to Labour et al. \cite{LGS01}, annealing of isotactic
polypropylene in the range of temperatures between 110 and 130 $^{\circ}$C
results in an increase in the concentration of ductile $\beta$ spherulites
that enhance plastic flow.
This phenomenon is explained by the fact that ``in the hexagonal phase, the chains
are loosely packed, which suggests that chains are mobile" \cite{HAR97}.
At annealing above 130 $^{\circ}$C, development of $\beta$ spherulites
is thermodynamically unfavorable (Al-Raheil et al. \cite{RQS98} reported that
``no $\beta$ spherulites were found above the crystallization temperature
132 $^{\circ}$C"), which implies an increase in $\epsilon_{\ast}$ 
and a corresponding decrease in elastoplastic deformations.

\item
Men and Strobl \cite{MS02} recently suggested that development of subsidiary lamellae
at annealing is a two-step process consisting of formation of lamellar blocks
at the initial stage and their aggregation into a ``blocky substructure"
at the final stage.
According to this scenario, enchancement of plastic flow at low-temperature annealing 
is attributed to sliding of isolated lamellar blocks formed at secondary crystallization
(the duration of thermal pre-treatment at temperatures in
the range of 110 to 130 $^{\circ}$C is assumed to be insufficient for the development of 
a blocky structure), whereas its decay at annealing above 130 $^{\circ}$C
is ascribed to aggregation of these blocks into stacks of relatively rigid lamellae.
\end{enumerate}
Both explanations are rather far from being exhausted.
The first is based on the assumption that $\beta$ spherulites grow
at low-temperature annealing, which was questioned in several works,
see, e.g., \cite{IS00} and the references therein.
The other explanation presumes a new mechanism for development of
subsidiary lamellae, which has not yet been confirmed experimentally.
It is worth also noting that these models do not establish links between
changes in the plastic strain, $\epsilon_{\ast}$, and appropriate
alternations in the rate of plastic flow, $b$.

A decrease in the rate of steady plastic flow, $b$, with temperature 
of annealing below 130 $^{\circ}$C is attributed to lamellar thickening, 
which results in slowing down of the developed plastic flow (which is associated
with lamellar fragmentation and motion of isolated lamellar blocks).
An increase in $b$ at higher temperatures is explained by the fact 
that at high-temperature annealing
two processes occur simultaneously: (i) thickening of dominant lamellae
and growth of subsidiary lamellae and (ii) annihilation of transverse lamellae.
Gu et al. \cite{GHT02} reported that annealing of iPP in the high-temperature
region leads to ``the complete absence of the cross-hatching."
This implies a substantial reduction in the rigidity of spherulites, and, 
as a consequence, an enhancement of the developed plastic flow.

To assess the level of elastoplastic strains at stretching of iPP specimens
annealed at different temperatures, we integrate Eqs. (16) and (17)
with adjustable parameters determined by matching experimental
data in tensile tests.
The results of numerical simulation are presented in Figure 11.
This figure reveals that the elastoplastic strain, $\epsilon_{\rm p}$,
is practically independent of annealing temperature
(although the stress--strain diagrams depicted in Figure 1
are strongly affected by thermal pre-treatment).

Figure 7 demonstrates that the average activation energy for
separation of strands from temporary nodes, $\Omega$, decreases with
stress, $\sigma$, for non-annealed samples 
and increases for specimens annealed at 140 $^{\circ}$C.
This qualitative difference in the dependence $\Omega(\sigma)$
for annealed and non-annealed iPP may be associated with
transformation of smectic mesophase into lamellar blocks
during thermal treatment.
In a stress-free non-annealed specimen, the concentration of
flocks of smectic mesophase (``arrays of chains with a better order
in longitudinal that in transverse chain direction" \cite{AW01})
is rather large.
These ordered arrays of macromolecules slow down rearrangement of strands
in active MRs,
which implies that at small stresses, the average activation energy of
samples not subjected to thermal treatment is rather large.
According to the Men--Strobl model \cite{MS02}, annealing of specimens 
results in transformation of smectic mesophase into lamellar blocks
and subsidiary lamellae.
This transformation of ``clusters of slow junctions" in amorphous regions
into the crystalline phase enhances the rearrangement process, 
which implies that at small stresses, $\Omega$ for
non-annealed specimens substantially exceeds that for annealed
samples.

With an increase in stress, clusters of smectic mesophase in
amorphous regions are disintegrated, and these ``ordered" chains
are transformed into ordinary ones.
This process enchances detachment of strands from
temporary junctions, which is reflected by a decrease in the average
potential energy of active meso-domains.

On the contrary, the growth of longitudinal stress results in fragmentation
of ``weak" lamellar blocks formed from smectic mesophase at annealing.
Pieces of broken lamellar blocks distributed in amorphous regions 
serve as extra physical cross-links with high activation energy 
for separation of strands.
As a result, stretching of annealed specimens increases
the potential energy for detachment of chains and slows down
the rearrangement process.

This scenario is confirmed by experimental data for the standard deviation
of potential energies of active meso-domains depicted in Figure 8.
This figure reveals that $\Sigma$ decreases with stress for
non-annealed specimens (which is attributed to destruction of
clusters of smectic mesophase in amorphous regions) and
increases with stress for annealed samples (which is ascribed to
the growth of concentration of extra physical cross-links induced
by breakage of ``weak" lamellar blocks).

For the quasi-Gaussian distribution function (5), 
$\Omega$ and $\Sigma$ may be treated as an apparent 
average potential energy for detachment of active strainds, 
and an apparent standard deviation of potential energies 
for separation of strands from the network.
The average activation energy, $\Omega_{0}$, 
and the standard deviation of activation energies, $\Sigma_{0}$, 
are given by
\[
\Omega_{0}=\int_{0}^{\infty} \omega p(\sigma^{0},\omega) d\omega,
\qquad
\Sigma_{0}=\biggl [ \int_{0}^{\infty} (\omega-\Omega_{0})^{2} p(\sigma^{0}, \omega) d\omega 
\biggr ]^{\frac{1}{2}}.
\]
We determine the dimensionless quantities $\Omega_{0}$ and $\Sigma_{0}$ 
according to these formulas and calculate the ratio
\[ 
\xi=\frac{\Sigma_{0}}{\Omega_{0}}
\]
that characterizes the width of the quasi-Gaussian distribution.
The parameter $\xi$ is plotted versus stress, $\sigma$, in Figure 11.
The experimental data are approximated by the linear function
\begin{equation}
\xi=\xi_{0}+\xi_{1} T,
\end{equation}
where the coefficients $\xi_{m}$ ($m=0,1$) are found by the least-squares
technique.
Figure 11 demonstrates that the width of the distribution of active MRs, Eq. (5),
is practically not affected by thermal treatment, and it weakly
increases with longitudinal stress.
The rate of increase in $\xi$ with stress appears to be independent of
annealing temperature.

Figure 9 reveals qualitatively different effects of the longitudinal stress
on the concentration of active meso-regions, $\kappa$, for non-annealed
specimens and for specimens subjected to thermal pre-treatment.
For non-annealed specimens, $\kappa$ remains constant up to 
$\sigma_{\ast} \approx 18$ MPa and linearly decreases with stress at higher
longitudinal stresses.
For specimens, annealed at 140 $^{\circ}$C, $\kappa$ slightly increases
with stress in the entire interval of deformations under consideration.

At relatively small stresses, $\sigma< \sigma_{\ast}$, the fraction of active
meso-domains in non-annealed specimens noticeably exceeds that
for annealed samples.
This conclusion is explained by the fact that annealing of iPP results
in secondary crystallization of a part of the amorphous phase,
which implies that the content of active MRs is reduced.
An increase in $\kappa$ with stress for annealed specimens is
also quite natural, because it is associated with mechanically-induced
fragmentation of weak lamellae and release of the amorphous phase,
whose rearrangement was prevented by surrounding lamellae
in a stress-free specimen.

To explain a pronounced decrease in $\kappa$ for non-annealed specimens
at stresses exceeding $\sigma_{\ast}$, we should recall that the fitting 
procedure presumed the quantities $\Omega$, $\Sigma$ and 
$\kappa$ in Eqs. (5) and (30) to be uniquelly determined by stress, $\sigma$, 
and to be independent of elastoplastic strain, $\epsilon_{\rm p}$.
To assess, whether this hypothesis is adequate, we integrate numerically
Eqs. (2), (5), (30), (32) and (33) with the adjustable parameters found by
matching observations and calculate the elastoplastic strain, $\epsilon_{\rm p}$,
as a function of time, $t$.
The results of numerical simulation are depicted in Figures 13 and 14.

Figure 14 shows that for all stresses (except for the highest stress, $\sigma=30$ MPa),
the elastoplastic strain, $\epsilon_{\rm p}$, slightly increases with time in a fashion
typical of primary creep.
Curve 5 in this figure demonstrates a transition from the primary creep to the
secondary creep (a linear increase in $\epsilon_{\rm p}$ with time) at the final
stage of the creep test.

In contrast with results presented in Figure 14, 
Figure 13 reveals that $\epsilon_{\rm p}$ increases with time following the 
primary-creep mode only at relatively small stresses ($\sigma<\sigma_{\ast}$).
At higher stresses, the primary creep is transformed into the secondary creep
(curves 4 and 5), and, finally, into the ternary creep (curve 6).

Based on this observation, one can conclude that below the critical stress, $\sigma_{\ast}$,
the concentration of active MRs, $\kappa$, is independent of the elastoplastic
strain, $\epsilon_{\rm p}$.
This implies that in the interval of stresses $[0,\sigma_{\ast}]$, 
$\kappa$ remains constant (in agreement with the experimental data 
plotted in Figure 9).
The independence of $\kappa$  of longitudinal stress, $\sigma$, in this
interval may be ascribed to the fact that at relatively small deformations
smectic mesophase is mainly rearranged, while dominant lamellae (that restrict
mobility of amorphous meso-domains) remain undisturbed.

Above the critical stress, $\sigma_{\ast}$, i.e., in the region of secondary
and ternary creep flows, the parameter $\kappa$ becomes a function of two
arguments: the longitudinal stress, $\sigma$, and the elastoplastic strain, $\epsilon_{\rm p}$.
This implies that the data for $\kappa$ depicted in Figure 9 at $\sigma>\sigma_{\ast}$
should be treated as ``average" (over the creep curves) quantities.
This means that the noticeable decrease in $\kappa$ (curve 1b in Figure 9) should
be attributed to the influence of secondary and ternary creep flows on rearrangement 
of strands in active MRs.

The following scenario may be proposed for a decrease in $\kappa$ observed
in creep tests at relatively large stresses.
A pronounced growth of elastoplastic strain $\epsilon_{\rm p}$ (curves 4 to 6 in
Figure 13) is associated with fragmentation of dominant lamellae
and plastic flow of lamellar blocks.
In contrast with annealed specimens, where primary lamellae are broken
into small pieces that serve as extra physical cross-links in active MRs 
(curves 2 in Figures 7 and 8), dominant lamellae in specimens not subjected
to thermal treatment are assumed to be disintegrated into 
relatively large blocks whose average length is comparable 
with the characteristic size of active MRs.
Plastic flow of these blocks results in confining of some amorphous regions,
where rearrangement of strands becomes prevented by surrounding
immobile lamellae and moving lamellar blocks.
This confinement of active MRs is temporary: when a lamellar block
moves away from a ``trapped" meso-region, it is released, and the rearrangement
process proceeds. 
However, the higher the plastic strain, $\epsilon_{\rm p}$, is, the large is the number of
``moving lamellar blocks," and, as a consequence, the smaller is the concentration
of active MRs.

According to this picture, an increase in the content of active meso-domains
with stress for annealed samples is attributed to breakage of lamellae into 
small blocks, whereas a decrease in $\kappa$ with stress for non-annealed
specimens is ascribed to fragmentation of lamellae into large blocks,
whose size is comparable with that of active MRs.
This difference in the processes of lamellar disintegration
for annealed and non-annealed samples is in agreement with 
the micro-mechanisms of lamellar growth at secondary crystallization 
recently proposed by Hikosaka et al. \cite{HAR97}.
According to it, lamellar thickening is accompanying by transformation of
folded chain crystals in the lamellar cores into extended chain crystals.
Packing of extended chains in the central part of a growing lamella 
becomes substantially looser compared to the initial one, which implies 
that dominant lamellae after thickening at secondary crystallization are fragmented into
noticeably smaller pieces than the same lamellae before annealing.

\section{Concluding remarks}

A model has been developed for the elastoplastic and viscoelastic
responses of semicrystalline polymers at isothermal loading.
A complicated micro-structure of a semi-crystalline polymer
is replaced by an equivalent transient network of macromolecules 
bridged by junctions (physical cross-links, entanglements and crystalline lamellae).
The network is thought of as an ensemble of meso-regions with various 
potential energies for separation of strands from temporary nodes.

The viscoelastic response of a semicrystalline polymer is attributed
to (i) detachment of active strands from temporary nodes in active 
meso-domains and (ii) merging of dangling strands with the network.
Rearrangement of strands is treated as a thermally activated process, 
whose rate is determined by the Eyring equation (2).

The elastoplastic behavior is ascribed to slippage of junctions 
with respect to their positions in the bulk material.
Kinetic equation (16) is proposed for the rate of sliding,
where the plastic flow is governed by the elastic strain
(not stress, as in conventional theories of plasticity).

Stress--strain relation (28) has been developed 
for isothermal uniaxial deformation
by using the laws of thermodynamics.
Adjustable parameters in the constitutive equations
are found by fitting experimental data in tensile tests with
a constant strain rate and in creep tests.

A series of tensile tests has been performed on 
isotactic polypropylene at room temperature.
The mechanical experiments are carried out on injection-molded 
specimens not subjected to thermal treatment 
and on samples annealed for 24 h at the temperatures
110, 120, 130, 140 and 150 $^{\circ}$C.

Two series of creep tests have been performed on non-annealed
specimens and on specimens annealed at 140 $^{\circ}$C
in the interval of stresses from 10 to 30 MPa.
Fair agreement is demonstrated between the experimental data
and the results of numerical simulation.

The following conclusions are drawn:
\begin{enumerate}
\item
The flow rule (15), (16) and (29) with 3 adjustable parameters
correctly describes observations in ``rapid" tensile tests, 
when the effect of material viscosity is negligible.
The elastic modulus, $E$, increases with annealing time
(which is attributed to lamellar thickening),
whereas the plastic strain, $\epsilon_{\ast}$, and the rate of
developed plastic flow, $b$, demonstrate a more sophisticated
behavior: they linearly decrease with annealing temperature, $T$,
below $T_{\rm c}=130$ $^{\circ}$C and increase above $T_{\rm c}$.
The temperature $T_{\rm c}$ roughly coincides with the temperature,
at which the rate of lamellar thickening noticeably increases.

\item
The governing equations (2), (5), (30), (32) and (33) with 3 material constants
adequately describe experimental data in creep tests under the
assumption that the quantities $\Omega$, $\Sigma$ and $\kappa$
are stress-dependent.
The average activation energy for separation of chains from temporary
nodes, $\Omega$, increases with stress for annealed specimens
and decreases for samples not subjected to thermal treatment.
This difference is attributed to different mechanisms of lamellar 
fragmentation into blocks.
The width of the quasi-Gaussian distribution (5) is weakly affected
by longitudinal stress and annealing temperature.

\item
The concentration of active meso-domains, $\kappa$, grows with stress
for annealed specimens.
This parameter is stress-independent for non-annealed samples below
the threshold stress, $\sigma_{\ast}$, and decreases with stress
at $\sigma>\sigma_{\ast}$.
The latter observation is ascribed to transition from the primary creep
of non-annealed specimens to the secondary and ternary creep flows,
when rearrangement of strands in active MRs becomes strongly affected by
elastoplastic strains.
\end{enumerate}

\newpage

\section*{List of figures}
\parindent 0 mm

{\bf Figure 1:}
The stress $\sigma$ MPa versus strain $\epsilon$ in tensile tests.
Circles: experimental data.
Curve 1: a non-annealed specimen;
curves 2 to 6: specimens annealed at a temperature $T$ $^{\circ}$C.
Curve 2: $T=110$;
curve 3: $T=120$;
curve 4: $T=130$;
curve 5: $T=140$;
curve 6: $T=150$.
Solid lines: results of numerical simulation
\vspace*{1 mm}

{\bf Figure 2:}
The strain $\epsilon$ versus time $t$ s in a tensile 
creep test with a stress $\sigma$ MPa.
Circles: experimental data for non-annealed specimens.
Solid lines: results of numerical simulation.
Curve 1:  $\sigma=10.0$;
curve 2:  $\sigma=15.0$;
curve 3:  $\sigma=17.5$;
curve 4:  $\sigma=20.0$;
curve 5:  $\sigma=22.5$;
curve 6:  $\sigma=25.0$
\vspace*{1 mm}

{\bf Figure 3:}
The strain $\epsilon$ versus time $t$ s in a tensile 
creep test with a stress $\sigma$ MPa.
Circles: experimental data for specimens annealed 
at $T=140$ $^{\circ}$C.
Solid lines: results of numerical simulation.
Curve 1:  $\sigma=10.0$;
curve 2:  $\sigma=15.0$;
curve 3:  $\sigma=20.0$;
curve 4:  $\sigma=25.0$;
curve 5:  $\sigma=30.0$
\vspace*{1 mm}

{\bf Figure 4:}
The elastic modulus $E$ GPa versus annealing temperature $T$ $^{\circ}$C.
Circles: treatment of observations.
Solid lines: approximation of the experimental data by Eq. (34).
Curve 1: $E_{0}=2.0242$, $E_{1}=0.0010$;
curve 2: $E_{0}=0.4946$, $E_{1}=0.0129$
\vspace*{1 mm}

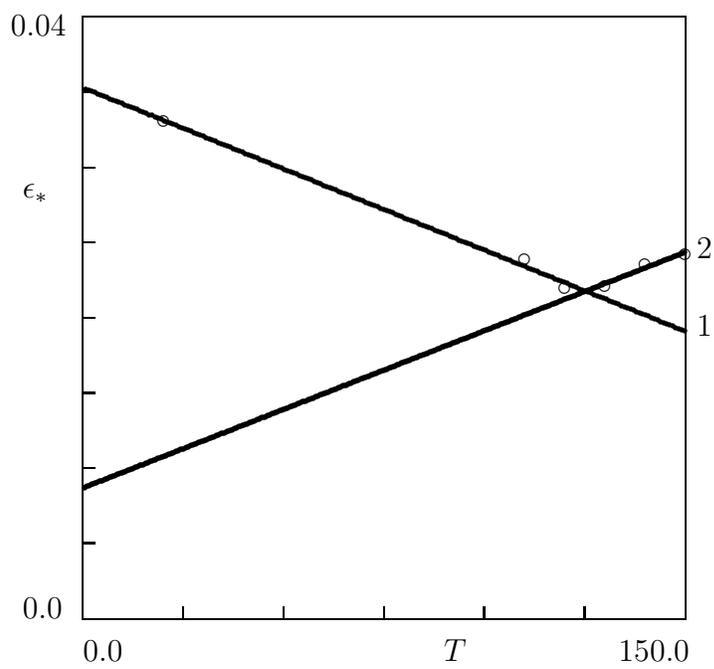
{\bf Figure 5:}
The plastic strain $\epsilon_{\ast}$ versus annealing temperature $T$ $^{\circ}$C.
Circles: treatment of observations.
Solid lines: approximation of the experimental data by Eq. (34).
Curve 1: $\epsilon_{\ast 0}=3.53\cdot 10^{-2}$, 
$\epsilon_{\ast 1}=-1.08\cdot 10^{-4}$;
curve 2: $\epsilon_{\ast 0}=8.68\cdot 10^{-3}$, 
$\epsilon_{\ast 1}=1.04\cdot 10^{-4}$
\vspace*{1 mm}

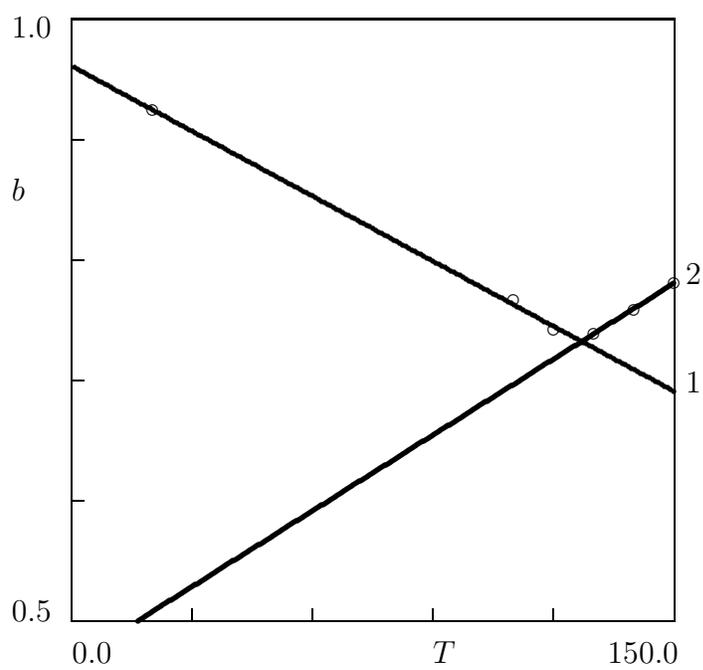
{\bf Figure 6:}
The rate of developed plastic flow $b$ versus annealing temperature $T$ $^{\circ}$C.
Circles: treatment of observations.
Solid lines: approximation of the experimental data by Eq. (34).
Curve 1: $b_{0}=0.9614$, $b_{1}=-0.0018$;
curve 2: $b_{0}=0.4657$, $b_{1}=0.0021$
\vspace*{1 mm}

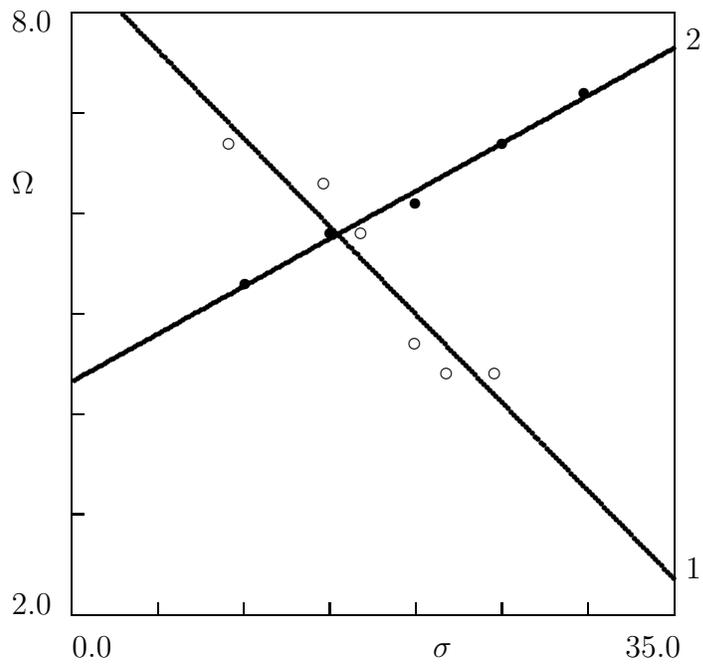
{\bf Figure 7:}
The average potential energy for detachment of strands 
$\Omega$ versus stress $\sigma$ MPa.
Symbols: treatment of observations.
Unfilled circles: non-annealed specimens;
filled circles:  specimens annealed at $T=140$ $^{\circ}$C.
Solid lines: approximation of the experimental data by Eq. (35).
Curve 1: $\Omega_{0}=8.5079$, $\Omega_{1}=-0.1756$;
curve 2: $\Omega_{0}=4.3232$, $\Omega_{1}=0.0950$
\vspace*{1 mm}

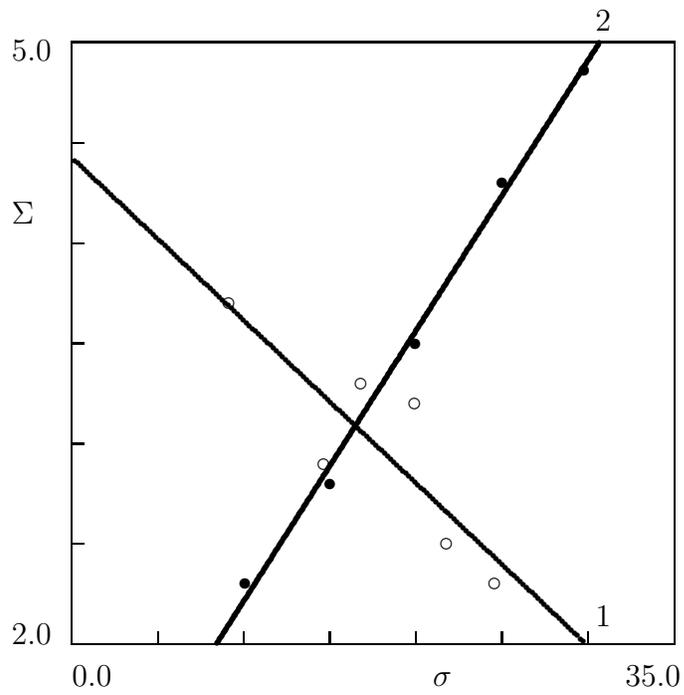
{\bf Figure 8:}
The standard deviation of potential energies for separation of strands 
$\Sigma$ versus stress $\sigma$ MPa.
Symbols: treatment of observations.
Unfilled circles: non-annealed specimens;
filled circles:  specimens annealed at $T=140$ $^{\circ}$C.
Solid lines: approximation of the experimental data by Eq. (35).
Curve 1: $\Sigma_{0}=4.4272$, $\Sigma_{1}=-0.0811$;
curve 2: $\Sigma_{0}=0.8672$, $\Sigma_{1}=0.1347$
\vspace*{1 mm}

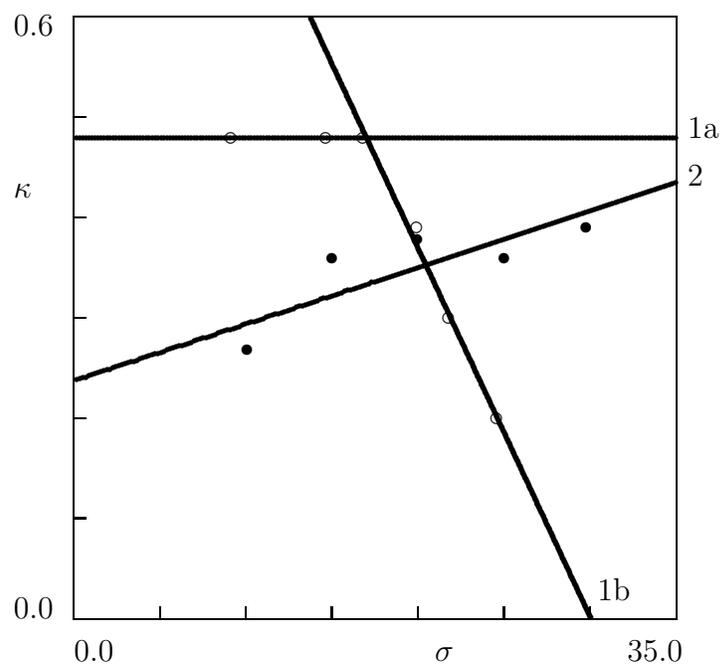
{\bf Figure 9:}
The concentration of active MRs $\kappa$ versus stress $\sigma$ MPa.
Symbols: treatment of observations.
Unfilled circles: non-annealed specimens;
filled circles:  specimens annealed at $T=140$ $^{\circ}$C.
Solid lines: approximation of the experimental data by Eq. (35).
Curve 1a: $\kappa_{0}=0.48$, $\kappa_{1}=0.0$;
curve 1b: $\kappa_{0}=1.1052$, $\kappa_{1}=-0.0367$;
curve 2: $\kappa_{0}=0.2371$, $\kappa_{1}=0.0057$
\vspace*{1 mm}

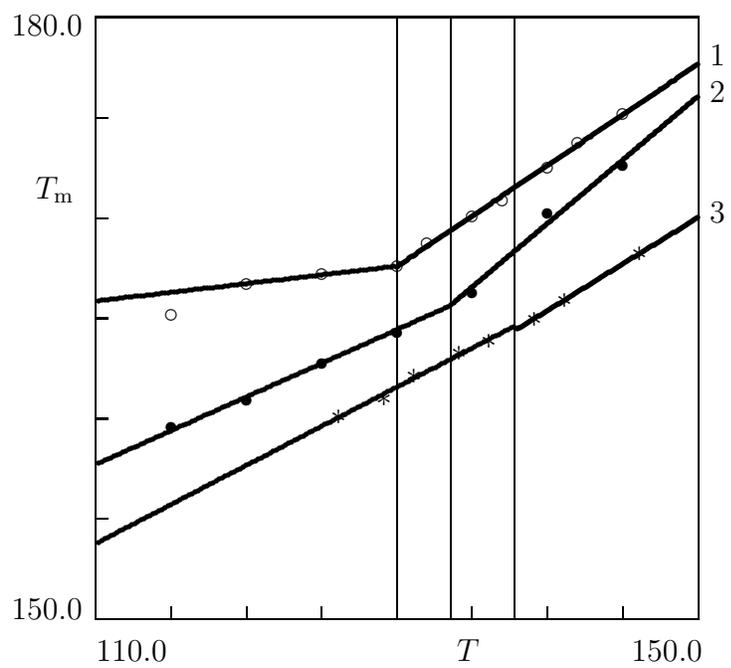
{\bf Figure 10:}
The melting peak $T_{\rm m}$ $^{\circ}$C versus crystallization
temperature $T$ $^{\circ}$C.
Symbols: treatment of observations for isotactic polypropylene.
Unfilled circles: iPP crystallized for 22 h at temperature $T$ 
and cooled to room temperature \cite{RQS98};
filled circles: iPP crystallized for an unspecified time 
at temperature $T$ \cite{IS00};
asterisks:  iPP crystallized for 100 min 
at temperature $T$ \cite{XSM98}.
Solid lines: approximation of the experimental data by Eq. (36).
Vertical lines indicate transition temperatures
\vspace*{1 mm}

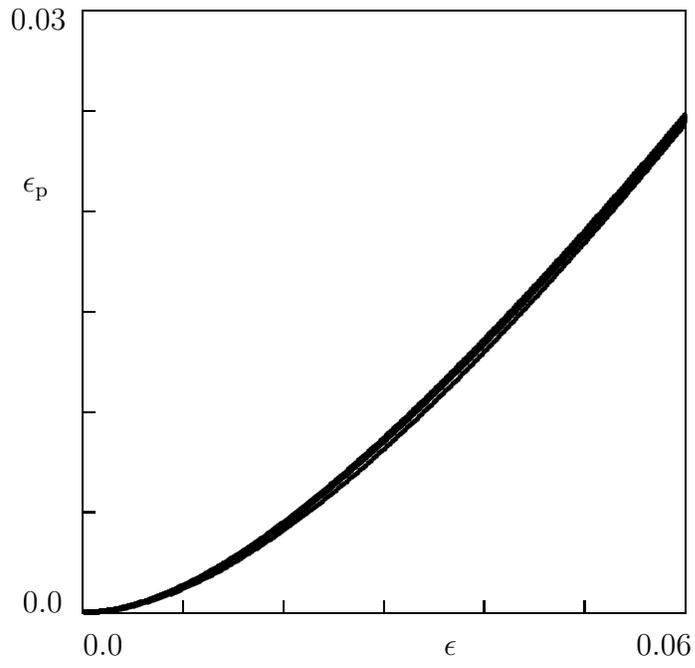
{\bf Figure 11:}
The elastoplastic strain $\epsilon_{\rm p}$ versus strain $\epsilon$ 
in tensile tests on specimens annealed at various temperatures $T$. 
Solid lines: results of numerical simulation
\vspace*{1 mm}

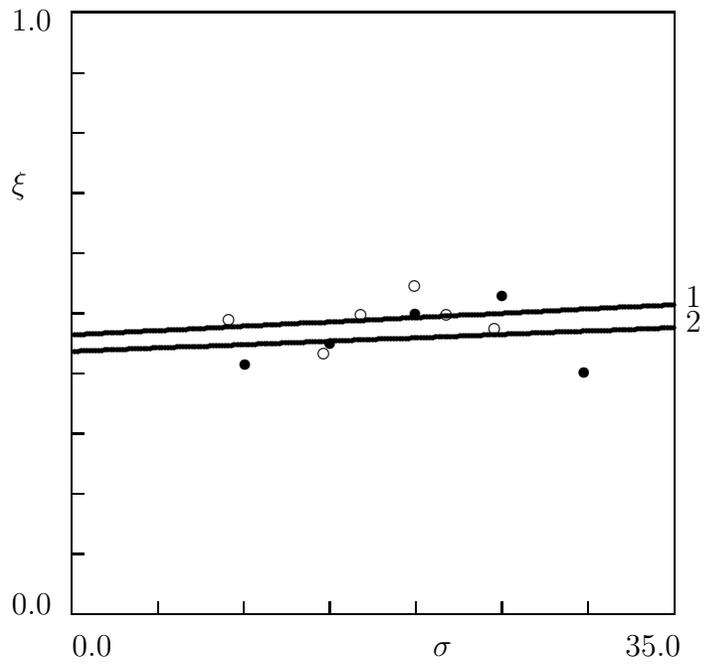
{\bf Figure 12:}
The ratio $\xi$ versus stress $\sigma$ MPa.
Symbols: treatment of observations.
Unfilled circles: non-annealed specimens;
filled circles:  specimens annealed at $T=140$ $^{\circ}$C.
Solid lines: approximation of the experimental data by Eq. (37).
Curve 1: $\xi_{0}=0.4643$, $\xi_{1}=0.0014$;
curve 2: $\xi_{0}=0.4369$, $\xi_{1}=0.0011$
\vspace*{1 mm}

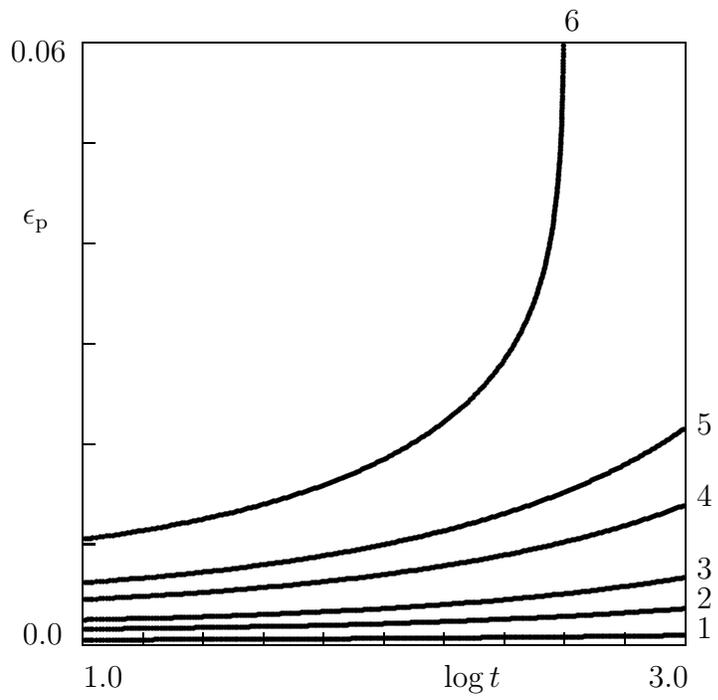
{\bf Figure 13:}
The elastoplastic strain $\epsilon_{\rm p}$ versus time $t$ s 
in a tensile creep test with a stress $\sigma$ MPa.
Solid lines: results of numerical simulation for non-annealed specimens.
Curve 1: $\sigma=10.0$;
curve 2: $\sigma=15.0$;
curve 3: $\sigma=17.5$;
curve 4: $\sigma=20.0$;
curve 5: $\sigma=22.5$;
curve 6: $\sigma=25.0$
\vspace*{1 mm}

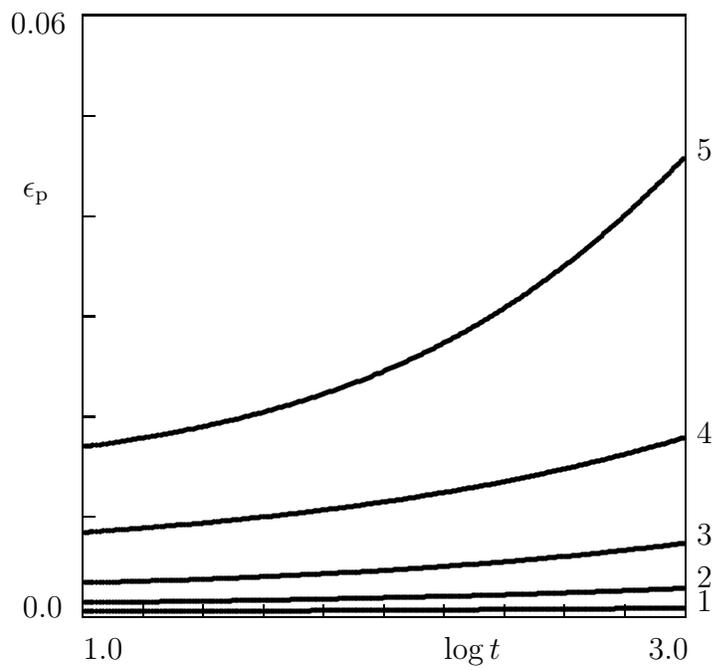
{\bf Figure 14:}
The elastoplastic strain $\epsilon_{\rm p}$ versus time $t$ s 
in a tensile creep test with a stress $\sigma$ MPa.
Solid lines: results of numerical simulation for specimens 
annealed at 140 $^{\circ}$C.
Curve 1: $\sigma=10.0$;
curve 2: $\sigma=15.0$;
curve 3: $\sigma=20.0$;
curve 4: $\sigma=25.0$;
curve 5: $\sigma=30.0$
\vspace*{100 mm}

\setlength{\unitlength}{0.8 mm}
\begin{figure}[tbh]
\begin{center}

\end{center}
\vspace*{10 mm}

\caption{}
\end{figure}

\begin{figure}[tbh]
\begin{center}
%
\end{center}
\vspace*{10 mm}

\caption{}
\end{figure}

\begin{figure}[tbh]
\begin{center}
%
\end{center}
\vspace*{10 mm}

\caption{}
\end{figure}

\begin{figure}[tbh]
\begin{center}
%
\end{center}
\vspace*{10 mm}

\caption{}
\end{figure}

\newpage
\begin{thebibliography}{50}

\bibitem{KB97}
G. Kalay, M.J. Bevis,
J. Polym. Sci. B: Polym. Phys. 35 (1997) 241, 265.

\bibitem{RQS98}
I.A. Al-Raheil, A.M. Qudah, M. Al-Share,
J. Appl. Polym. Sci. 67 (1998) 1259, 1267.

\bibitem{XSM98}
J. Xu, S. Srinivas, H. Marand, P. Agarwal,
Macromolecules 31 (1998) 8230.

\bibitem{ABM99}
R.G. Alamo, G.M. Brown, L. Mandelkern, A. Lehtinen, R. Paukkerri,
Polymer 40 (1999) 3933.

\bibitem{IS00}
M. Iijima, G. Strobl,
Macromolecules 33 (2000) 5204.

\bibitem{MHY00}
P. Maiti, M. Hikosaka, K. Yamada, A. Toda, F. Gu,
Macromolecules 33 (2000) 9069.

\bibitem{GHT02}
F. Gu, M. Hikosaka, A. Toda, S.K. Ghosh, S. Yamazaki, 
M. Arakaki, K. Yamada,
Polymer 43 (2002) 1473.

\bibitem{WW66}
I.M. Ward, J.M. Wolfe, 
J. Mech. Phys. Solids 14 (1966) 131.

\bibitem{SW72}
J. Smart, J.G. Williams, 
J. Mech. Phys. Solids 20 (1972) 313.

\bibitem{WS94}
F.-J. Wortmann, K.V. Schulz, 
Polymer 35 (1994) 2108.

\bibitem{WS95}
F.-J. Wortmann, K.V. Schulz, 
Polymer 36 (1995) 2363.

\bibitem{Ari96}
T. Ariyama, 
J. Mater. Sci. 31 (1996) 4127.

\bibitem{AMK97}
T. Ariyama, Y. Mori, K. Kaneko,
Polym. Eng. Sci. 37 (1997) 81.

\bibitem{DE97}
N.K. Dutta, G.H. Edward, 
J. Appl. Polym. Sci. 66 (1997) 1101.

\bibitem{RT97}
B.E. Read, P.E. Tomlins,
Polymer 38 (1997) 4617.

\bibitem{TR98}
P.E. Tomlins, B.E. Read,
Polymer 39 (1998) 355.

\bibitem{SCC99}
J. Sweeney, T.L.D. Collins, P.D. Coates, R.A. Duckett, 
J. Appl. Polym. Sci. 72 (1999) 563.

\bibitem{AGU95}
M. Aboulfaraj, C. G'Sell, B. Ulrich, A. Dahoun,
Polymer 36 (1995) 731.

\bibitem{CCG98}
G. Coulon, G. Castelein, C. G'Sell, 
Polymer 40 (1998) 95.

\bibitem{SeSE99}
R. Seguela, E. Staniek, B. Escaig, B. Fillon,
J. Appl. Polym. Sci. 71 (1999) 1873.

\bibitem{StSE99}
E. Staniek, R. Seguela, B. Escaig, P. Francois,
J. Appl. Polym. Sci. 72 (1999) 1241.

\bibitem{NT99}
K.-H. Nitta, M. Takayanagi, 
J. Polym. Sci. B: Polym. Phys. 37 (1999) 357.

\bibitem{NT00}
K.-H. Nitta, M. Takayanagi, 
J. Polym. Sci. B: Polym. Phys. 38 (2000) 1037.

\bibitem{LGS01}
T. Labour, C. Gauthier, R. Seguela, G. Vigier, Y. Bomal, G. Orange,
Polymer 42 (2001) 7127.

\bibitem{VMH96}
R. Verma, H. Marand, B. Hsiao,
Macromolecules 29 (1996) 7767.

\bibitem{ZBC99}
X.C. Zhang, M.F. Butler, R.E. Cameron,
Polym. Int. 48 (1999) 1173.

\bibitem{MP01}
R.W. Meyer, L.A. Pruitt, 
Polymer 42 (2001) 5293.

\bibitem{SW96}
J. Sweeney, I.M. Ward, 
J. Mech. Phys. Solids 44 (1996) 1033.

\bibitem{And99}
E. Andreassen, 
Polymer 40 (1999) 3909.

\bibitem{TE92}
F. Tanaka, S.F. Edwards, 
Macromolecules 25 (1992) 1516.

\bibitem{ABJ95}
E.M. Arruda, M.C. Boyce, R. Jayachandran, 
Mech. Mater. 19 (1995) 193.

\bibitem{KE75}
A.S. Krausz, H. Eyring, 
Deformation Kinetics.
Wiley, New York (1975).

\bibitem{MS02}
Y. Men, G. Strobl, Polymer, 43 (2002) 2761.

\bibitem{HAR97}
M. Hikosaka, K. Amano, S. Rastogi, A. Keller,
Macromolecules 30 (1997) 2067.

\bibitem{AW01}
R. Androsch, B. Wunderlich,
Macromolecules 34 (2001) 5950.
\end{thebibliography}
\end{document}